\documentclass{article}
\usepackage{spconf}
\usepackage{algorithmic,amsmath,graphicx,hyperref,cite,amssymb}
\usepackage{textcomp}
\usepackage{xcolor}
\usepackage{bm}

\usepackage{booktabs}

\usepackage{adjustbox}
\usepackage{makecell}
\usepackage{hyperref}

\newcommand{\etal}{\textit{et al.}}
\usepackage{xr}
\title{Title of the paper}

\title{EMG-to-Speech with Fewer Channels}

\name{Injune Hwang$^{\star}$ \qquad Jaejun Lee$^{\star}$ \qquad Kyogu Lee$^{\star \dagger \ddagger}$}
\address{$^{\star}$Department of Intelligence and Information, Seoul National University, Republic of Korea\\ $^{\dagger}$Artificial Intelligence Institute, Seoul National University, Republic of Korea\\
$^{\ddagger}$Interdisciplinary Program in Artificial Intelligence, Seoul National University, Republic of Korea}
\begin{document}
\ninept
\maketitle
\begin{abstract}
Surface electromyography (EMG) is a promising modality for silent speech interfaces, but its effectiveness depends heavily on sensor placement and channel availability. In this work, we investigate the contribution of individual and combined EMG channels to speech reconstruction performance.
Our findings reveal that while certain EMG channels are individually more informative, the highest performance arises from subsets that leverage complementary relationships among channels. We also analyzed phoneme classification accuracy under channel ablations and observed interpretable patterns reflecting the anatomical roles of the underlying muscles. To address performance degradation from channel reduction, we pretrained models on full 8-channel data using random channel dropout and fine-tuned them on reduced-channel subsets. Fine-tuning consistently outperformed training from scratch for 4–6 channel settings, with the best dropout strategy depending on the number of channels. These results suggest that performance degradation from sensor reduction can be mitigated through pretraining and channel-aware design, supporting the development of lightweight and practical EMG-based silent speech systems.
\end{abstract}
\section{Introduction}
\label{sec:intro}
Electromyography (EMG) has emerged as a promising modality for silent speech interfaces, enabling speech decoding by measuring facial and neck muscle activity without relying on acoustic signals. Such an approach offers communication opportunities for individuals who are unable to generate voice using their vocal folds.

EMG-to-speech techniques have achieved notable advances with the advent of deep learning. Gaddy~\etal~\cite{gaddy2022voicing} introduced the first widely used EMG dataset and proposed a single-speaker EMG-to-speech synthesis system. Building upon this dataset, subsequent studies have explored directions such as EMG-to-text recognition~\cite{hou2024emg2vec} and extensions to multi-speaker synthesis~\cite{scheck2023multi,lee2025speaking}.

Although these studies have achieved significant academic progress, several challenges remain before practical deployment can be realized. In particular, conventional research relies on full-channel EMG sensor arrays that cover nearly half of the face. From the perspective of usability and accessibility, it is crucial to develop lightweight and user-friendly EMG-to-speech devices. Thus, maintaining high recognition performance with a reduced number of channels is a critical objective.

In this study, we systematically investigate channel efficiency in EMG-to-speech using the widely adopted dataset introduced by Gaddy~\etal~\cite{gaddy2022voicing}. We first perform a greedy analysis by iteratively removing one channel at a time to examine performance degradation, and further evaluate sub-channel combinations to assess the relative importance of each channel. To gain deeper insight, we analyze phoneme-level performance variations to estimate the articulatory contributions of individual channels. Moreover, we demonstrate that fine-tuning a model pretrained on full-channel data with reduced-channel inputs consistently outperforms training from scratch. 

Finally, we demonstrate that incorporating randomized channel dropout during pretraining yields additional performance gains under reduced-channel settings. This strategy further boosts generalization, particularly when fine-tuning on limited-channel inputs.

To contextualize, we delineate our contributions as follows:
\begin{itemize}
  \item We present a comprehensive analysis of EMG channel importance using greedy elimination and exhaustive subset evaluation, identifying both dominant and complementary channel contributions.
  \item We conduct phoneme-level ablation analyses across 7-channel settings to reveal how specific channels influence the recognition of distinct phoneme categories (e.g., fricatives, plosives).
  \item We propose a fine-tuning strategy that incorporates randomized channel dropout during pretraining, consistently improving performance in reduced-channel scenarios compared to training from scratch.
\end{itemize}

\section{Related work}
\subsection{EMG-based Speech Synthesis}
A widely used EMG dataset was introduced by Gaddy~\etal~\cite{gaddy2020digital}, which consists of eight-channel recordings of both voiced EMG (captured during audible speech) and silent EMG (captured during non-audible articulation) from a single speaker. Since parallel acoustic data is available only for the voiced EMG, they proposed a training strategy that leverages a Dynamic Time Warping (DTW) loss to align silent EMG with the corresponding voiced speech. By combining this alignment with a transformer-based EMG encoder that predicts mel-spectrograms subsequently rendered by a pretrained vocoder, their method achieved promising results in synthesizing speech from silent EMG. Building on this framework, Scheck~\etal~\cite{scheck2023multi} extended the approach to a multi-speaker setting by adopting a voice conversion framework~\cite{van2022comparison} that uses HuBERT~\cite{hsu2021hubert} representations as encoder targets. While this extension demonstrated strong performance in multi-speaker scenarios, thanks to its disentanglement capability, Gaddy’s mel-spectrogram–based method remains superior in single-speaker settings.

Notably, the Gaddy dataset has become a standard benchmark in the silent speech community and has been adopted in several recent studies. For example, Ren~\etal~\cite{ren2024diff} introduced a diffusion-based EMG-to-speech model, and Ullah~\etal~\cite{ullah2024optimized} proposed an optimized training framework tailored to silent EMG. In addition, Wu~\etal~\cite{wu2024deep} utilized the dataset to explore deep prompt tuning for personalized silent speech decoding, while Benster~\etal~\cite{benster2024cross} leveraged it to investigate cross-modal representations between EMG and acoustic features. These works collectively underscore the continued relevance and utility of Gaddy’s dataset for advancing silent speech synthesis research.

\subsection{Synthesis on reduced EMG channels}
Several studies have explored the impact of channel reduction on EMG-to-speech performance. Gaddy~\etal~\cite{gaddy2022voicing} investigated 35 randomly chosen four-channel subsets—half of the 70 possible combinations—providing partial insight into channel importance. Lee~\etal~\cite{lee2025articulatory} selected two to four channels based on their correlation with articulatory features, reporting improved performance; however, broader empirical validation is still required. While these works highlight the feasibility of EMG channel reduction, their findings remain limited in scope and insufficiently validated. In contrast, our work provides a more systematic and extensive investigation into channel efficiency and phoneme-level contributions.

\begin{figure}
    \centering
    \includegraphics[width=0.95\linewidth]{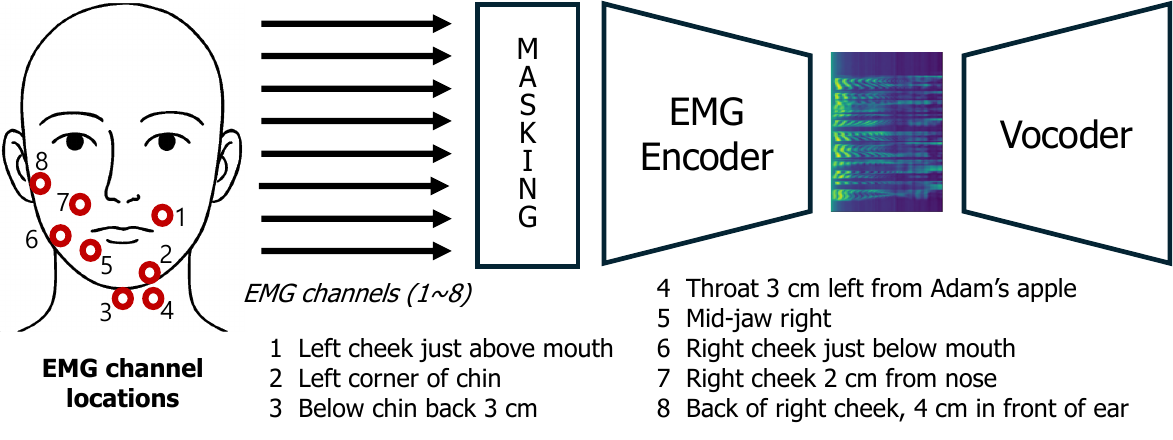}
    \caption{Overview of the EMG-to-speech framework. The masking block appears only in our fine-tuning variants. EMG channel locations are illustrated in accordance with the description below.}
    \label{fig:framework}
    \vspace{-1em}
\end{figure}

\section{Methods}
\subsection{Analysis under Reduced EMG Channel Conditions}
We hypothesize that not all EMG channels contribute equally to speech synthesis. Instead, mutual relationships may exist among channels, suggesting interdependencies and redundancies in their contributions. This implies that some channels are preferable to retain first when channel reduction is required, or that certain subsets may yield better performance when the number of channels is limited. To investigate this, we utilized the widely adopted eight-channel EMG dataset~\cite{gaddy2020digital} and its associated synthesis framework~\cite{gaddy2021improved}, which demonstrates strong performance in single-speaker settings. The overall framework is depicted on Figure~\ref{fig:framework}.
To estimate the relative contribution of each EMG channel to synthesis performance, we designed three complementary experiments that assess channel importance from different perspectives.

\subsubsection{Backward Elimination}
We investigated the effect of gradually reducing the number of channels to evaluate the relative contribution of each channel and to identify any performance drop-off threshold. Rather than exhaustively evaluating all possible channel combinations, we employed a greedy backward elimination strategy: starting from the full 8-channel set, we removed one channel at a time to form 8 candidate subsets, selected the best-performing subset, and repeated the process iteratively. Although this method does not account for combinatorial interactions or synergistic effects between channels, it provides a fast approximation of individual channel importance.

\subsubsection{Exhaustive Evaluation of 4-Channel Combinations}
We evaluated the performance of all possible 4-channel combinations out of the 8 available channels (a total of 70 combinations). This allowed us to assess the relative performance of the 4-channel subset identified by the greedy method in the first experiment and to compute the average WER for each channel across all combinations. By analyzing the frequency with which certain channels appeared in top-performing combinations, we also gained insight into their relative importance and complementary relationships. This exhaustive experiment enables direct comparison with prior works, such as the 35 random 4-channel subsets tested in Gaddy et al. \cite{gaddy2022voicing} and the articulatory-motivated 4-channel selection proposed in Lee et al. \cite{lee2025articulatory}.

\subsubsection{Phoneme Error Analysis on 7-Channel Variants}
We analyzed phoneme classification performance when each of the eight channels was removed individually (i.e., for all 7-channel subsets). This analysis aimed to assess how each channel affects recognition of different phoneme categories. To isolate the effect of EMG encoding alone, we bypassed the speech synthesis and ASR stages and instead evaluated phoneme predictions directly from the output of the EMG encoder. For deeper insight, phonemes were grouped by linguistic and articulatory features: consonants vs. vowels, voiced vs. voiceless, consonants by manner and place of articulation, and vowels by tongue height and backness. This categorization enabled examination of how individual channels contribute to fine-grained phonetic distinctions.

\begin{figure}
    \centering
    \includegraphics[width=0.8\linewidth]{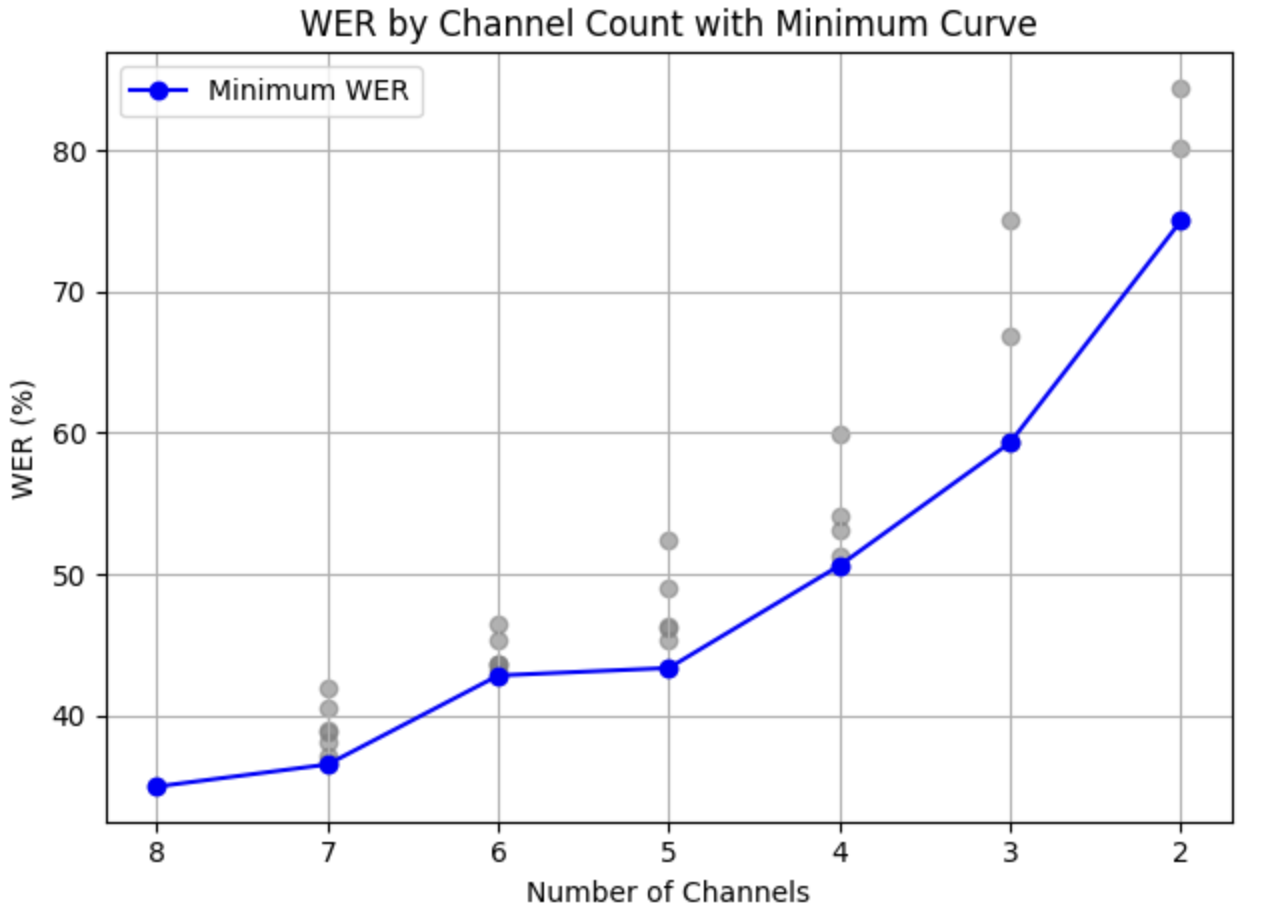}
    \caption{WERs of models using different numbers of EMG channels. Each dot represents a model configuration, and the blue line connects the best-performing configuration at each channel count.}
    \label{fig:backward_elimination}
    \vspace{-1em}
\end{figure}

\subsection{Finetuning with Channel Dropout} \label{method:finetune}
To address the performance drop associated with reduced-channel conditions, we initialized models using weights from a pretrained model trained on all eight channels. The goal was to adapt this model to scenarios with fewer channels via fine-tuning.

During the 8-channel pretraining, we applied channel dropout (channel masking) to improve generalization to reduced-channel conditions (Figure~\ref{fig:framework}). Applying dropout to enhance robustness is a widely adopted strategy across domains~\cite{guo2023place, he2021locality}; similarly, we applied channel dropout to improve generalization to varying channel availability. For each training utterance, individual channels were randomly masked with a fixed probability, independently sampled per channel. Channel dropout was applied only during training, not at inference time.

Formally, let $\bm{X} \in \mathbb{R}^{C \times T}$ be the input EMG signal with $C$ channels and $T$ time steps. We define a dropout mask $\bm{m} \in \{0, 1\}^C$ such that:

\[
m_c \sim \text{Bernoulli}(1 - p) \quad \text{for } c = 1, \dots, C
\]

where $p$ is the dropout probability. This mask is applied independently to each channel in every forward pass. The masked input $\tilde{\bm{X}}$ is then computed via:

\[
\tilde{\bm{X}} = \bm{m} \otimes \bm{X}
\]

where $\otimes$ denotes channel-wise broadcasting of the mask $\bm{m}$ over the time dimension. We selected the dropout probabilities $p \in \{0, 0.125, 0.25\}$, corresponding to using all channels, or retaining on average 7 or 6 out of the 8 channels, respectively. These values were chosen to simulate realistic sensor failures while avoiding excessive information loss that may destabilize training.



\begin{table}[t]
    \footnotesize
    \centering
    \caption{Top-10 WERs(\%) Among 4-Channel subsets.}
    \label{tab:top10}
    \vspace{0.2cm}
    \begin{tabular}{c|cccccccc|cc}
    \toprule
    \textbf{Subset} & 1 & 2 & 3 & 4 & 5 & 6 & 7 & 8 & WER \\
    \midrule
    1356 & 1 & 0 & 1 & 0 & 1 & 1 & 0 & 0 & 47.2 \\
    2357 & 0 & 1 & 1 & 0 & 1 & 0 & 1 & 0 & 47.3 \\
    1346 & 1 & 0 & 1 & 1 & 0 & 1 & 0 & 0 & 47.7 \\
    1238 & 1 & 1 & 1 & 0 & 0 & 0 & 0 & 1 & 48.3 \\
    1235 & 1 & 1 & 1 & 0 & 1 & 0 & 0 & 0 & 48.4 \\
    2347 & 0 & 1 & 1 & 1 & 0 & 0 & 1 & 0 & 48.6 \\
    1236 & 1 & 1 & 1 & 0 & 0 & 1 & 0 & 0 & 48.8 \\
    1345 & 1 & 0 & 1 & 1 & 1 & 0 & 0 & 0 & 49.0 \\
    1245 & 1 & 1 & 0 & 1 & 1 & 0 & 0 & 0 & 49.6 \\
    2367 & 0 & 1 & 1 & 0 & 0 & 1 & 1 & 0 & 49.6 \\
    \midrule
    Count & 7 & 7 & 9 & 4 & 5 & 4 & 3 & 1 \\
    \bottomrule
    \end{tabular}\\
    \vspace{-0.5cm}
\end{table}

\section{Experiments}
\subsection{Dataset and Preprocessing}
We conducted our experiments using the publicly available EMG dataset released by Gaddy et al. \cite{gaddy2020digital}, which has become a widely used benchmark in silent speech research. The dataset contains approximately 20 hours of EMG recordings from a single subject, including 16 hours of open vocabulary data and 4 hours of closed vocabulary data. We excluded the closed-vocabulary portion, as our study focuses on open-set generalization and channel robustness in continuous speech settings. EMG signals were recorded using eight surface electrodes placed on the face and neck (see Figure~\ref{fig:framework}). Each utterance is paired with a synchronized audio waveform, and the dataset includes time-aligned phoneme and word-level transcriptions.
We followed the standard preprocessing pipeline provided in the official GitHub repository\footnote{\url{https://github.com/dgaddy/silent\_speech}}, using mel spectrograms as acoustic targets.

\subsection{Model Architecture and Training Objective}
We adopted the model architecture proposed by Gaddy~\etal~\cite{gaddy2021improved}, which combines convolutional and Transformer layers for sequence modeling. On top of the encoder output, two separate linear projection layers are used: one to predict mel spectrograms and the other to classify phonemes. For the training objective, we follow Gaddy’s settings, including mel spectrogram reconstruction loss and phoneme classification loss.
When training the model with fewer EMG channels, we kept the entire architecture unchanged except for adjusting the input channel size of the first convolutional layer to match the number of available channels.

The predicted mel spectrograms were converted into waveforms using HiFi-GAN~\cite{kong2020hifi}, then transcribed using the Whisper ASR model(medium)~\cite{radford2023robust}. The resulting transcriptions were compared with the ground-truth text to compute the word error rate (WER).

All experiments used our open-source codebase. \footnote{\url{https://github.com/SPJune/SS_by_Channel}}

\section{Results and Analysis}

\subsection{Backward Elimination} \label{result:backward}
We analyzed channel importance through a backward elimination approach, where one channel was removed at a time and the model performance was measured. As shown in Fig.~\ref{fig:backward_elimination}, channels were eliminated in the following order: channel 6, 7, 8, 5, 4, and 1, leaving channels 2 and 3 as the last retained and therefore most critical. When each channel was removed individually, the impact on WER (from most to least detrimental) followed the order: channel 3, 2, 7, 1, 5, 4, 8, and 6. When using only a single channel, the WER exceeded 100\%, rendering quantitative comparisons uninformative. 
Interestingly, some subsets with fewer channels outperformed larger subsets, suggesting that certain channels may contribute redundant or even noisy information that hinders generalization.
As the number of channels decreased, performance variance across different channel combinations increased, indicating heightened sensitivity to channel selection under reduced-channel settings. 

\begin{table}[t]
    \footnotesize
    \setlength{\tabcolsep}{5.5pt}
    \centering
    \caption{Channel-wise average WERs(\%) based on all 4-channel subsets that include each channel.}
    \label{tab:avg_wer_single_channels}
    \vspace{0.2cm}
    \begin{tabular}{c|cccccccc}
        \toprule
        \textbf{Ch.} & 3 & 2 & 1 & 5 & 6 & 4 & 7 & 8 \\
        \midrule
        \textbf{WER} & 51.4 & 52.3 & 52.6 & 52.8 & 53.1 & 53.7 & 53.8 & 54.8 \\
        \bottomrule
    \end{tabular}
    \vspace{-0.5cm}
\end{table}

\subsection{Exhaustive Evaluation of 4-Channel Combinations}

To complement the greedy backward elimination analysis, we conducted an exhaustive evaluation of all 70 possible 4-channel combinations. Among these, at least 10 combinations outperformed the greedy-selected subset (channels 1, 2, 3, and 4), as shown in Table~\ref{tab:top10}.
Notably, when channel 2 was absent, channels 5 or 6 were often selected as substitutes. Similarly, when channel 1 was excluded, channel 7 frequently appeared in the top-10 combinations. This suggests that certain channels carry complementary information and can compensate for each other depending on the combination.
Considering approximate spatial locations, channel 1 is positioned on the upper left of the face, 7 on the upper right, 2 on the lower left, and 5 and 6 on the lower right. This spatial diversity may contribute to capturing overlapping or complementary muscle activation regions.
Interestingly, channel 6—previously eliminated first in backward elimination—appeared in several top-performing combinations. Conversely, channel 2, which was considered most critical in the greedy analysis, was absent from some of the top-ranked combinations.
Based on the average WER across all combinations in which each channel appeared, the channels were ranked in the following order: 3, 2, 5, 1, 6, 4, 7, and 8 (Table~\ref{tab:avg_wer_single_channels}). Frequency analysis of top-10 subsets showed that channel 3 appeared most frequently, followed by channels 1 and 2, then 5, 4, 6, 7, and 8.
Finally, the best-performing subset reported in \cite{gaddy2022voicing} (channels 1, 2, 3, 8) ranked 4th in our results, and the articulatory-motivated subset from \cite{lee2025articulatory} (1, 2, 3, 5) ranked 5th. These findings highlight the value of conducting exhaustive subset evaluations to identify optimal configurations beyond what greedy or heuristic methods can capture.
\begin{table}[t]
    \footnotesize
    \setlength{\tabcolsep}{5pt}
    \caption{Error rate of each phoneme category for 8-channel baseline and the worst-performing single-channel ablation.}
    \centering
    \begin{tabular}{l|r|r|c}
    \toprule
    Category & 8 ch PER & worst PER & most critical ch. \\
    \midrule
    total PER & 16.0 & 17.1 & 8 \\
    \midrule
    vowel & 22.6 & 24.1 & 7 \\
    consonant & 20.9 & 22.7 & 3 \\
    silence & 4.5 & 5.8 & 8 \\
    \midrule
    voiced & 21.3 & 23.2 & 7 \\
    voiceless & 22.4 & 25.3 & 3 \\
    \midrule
    manner\_liquid & 12.7 & 16.9 & 1 \\
    manner\_fricative & 19.4 & 23.2 & 3 \\
    manner\_nasal & 21.0 & 23.5 & 7 \\
    manner\_plosive & 26.4 & 27.6 & 7 \\
    \midrule
    place\_bilabial & 21.6 & 25.2 & 8 \\
    place\_alveolar & 21.4 & 23.0 & 3 \\
    place\_labiodental & 17.9 & 22.5 & 2\\
    place\_velar & 18.5 & 21.5 & 1\\
    \midrule
    vowel\_high & 21.7 & 28.8 & 7 \\
    vowel\_mid & 23.7 & 25.1 & 7 \\
    vowel\_low & 22.6 & 22.7 & 3 \\
    \midrule
    vowel\_front & 24.4 & 28.8 & 7 \\
    vowel\_central & 24.9 & 25.7 & 8 \\
    vowel\_back & 17.7 & 20.7 & 6 \\
    \midrule
    vowel\_rounded & 17.8 & 20.7 & 6 \\
    vowel\_unrounded & 23.6 & 24.8 & 7 \\
    \bottomrule
    \end{tabular}
    \label{tab:phoneme_worst}
    \vspace{-0.5cm}
\end{table}

\subsection{Phoneme Error Analysis on 7-Channel Variants}

To examine how individual EMG channels contribute to different phoneme categories, we measured phoneme error rates (PER) across all 7-channel subsets, as summarized in Table~\ref{tab:phoneme_worst}. To ensure statistical reliability, phoneme categories with fewer than 3,000 samples (approximately 2\% of the corpus) were excluded. These include affricates, glides, postalveolars, glottals, labiovelars, and palatals. 
Channel 8 showed the highest overall PER increase when removed. Positioned over the posterior masseter, it likely captures jaw movement and lip closure. Its absence degraded recognition of bilabials, silence segments, and central vowels. Channel 7, near the zygomaticus major, affected high front vowel recognition, indicating its role in shaping cheek and oral cavity contours. Channel 3, located near the sternohyoid, influenced voiceless fricatives and low vowels, consistent with its involvement in tongue and larynx positioning. Channel 2 (depressor anguli oris) affected labiodental phonemes, while channel 6 (orbicularis oris) impacted round vowels.
The contribution of channel 1 to liquid and velar phoneme recognition remains unclear. Importantly, the absence of performance degradation upon removing certain channels does not necessarily imply irrelevance; rather, it may indicate signal redundancy or compensation by other sensors.
Interestingly, channels 7 and 8—deemed less critical in synthesis evaluations—played key roles in phoneme discrimination. This highlights a divergence in channel utility between speech synthesis and phoneme recognition tasks.

\begin{figure}
    \centering
    \includegraphics[width=0.8\linewidth]{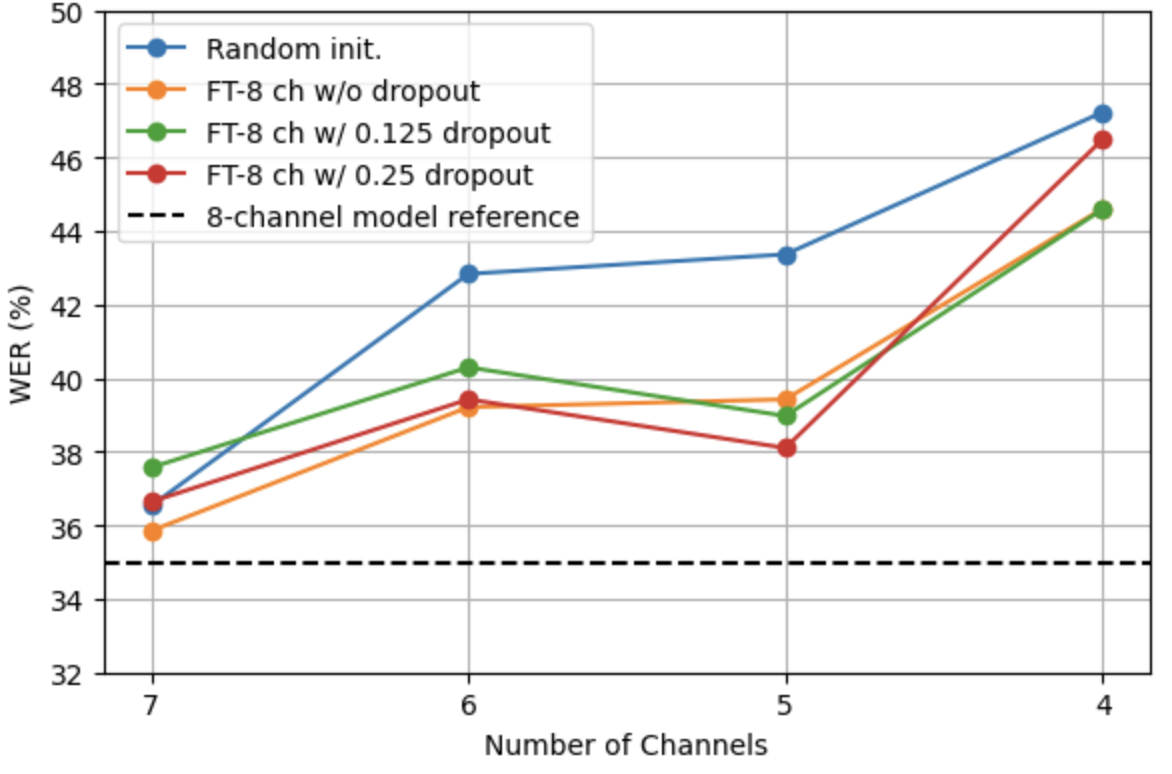}
    \caption{Effect of channel dropout and fine-tuning on WER across different channel configurations. 8-channel performance shown as reference.}
    \label{fig:finetuning}
    \vspace{-1em}
\end{figure}

\subsection{Fine-Tuning with Channel Dropout}

To mitigate performance degradation under reduced-channel conditions, we fine-tuned models initialized from an 8-channel pretrained baseline incorporating our proposed channel dropout strategy (Section~\ref{method:finetune}).
Fine-tuning was applied to the best-performing channel subsets: for 7 channels, (1, 2, 3, 4, 5, 7, 8); for 6 channels, (1, 2, 3, 4, 5, 8); for 5 channels, (1, 2, 3, 4, 5); and for 4 channels, (1, 3, 5, 6). The 4-channel subset was selected based on the exhaustive search results, while others were based on backward elimination.
As shown in Fig.~\ref{fig:finetuning}, performance in the 7-channel setting was comparable across initialization methods. In contrast, for 4, 5, and 6 channels, fine-tuned models consistently outperformed those trained from scratch, demonstrating the utility of full-channel pretraining even when deploying reduced-channel models.
Interestingly, the best-performing dropout setting varied by channel count: for 6 and 4 channels, no dropout ($p = 0$) yielded the best results, while for 5 channels, higher dropout improved performance. Remarkably, although 5-channel models underperformed 6-channel models when trained from scratch, the trend reversed after fine-tuning. This suggests that model performance is not solely determined by the number of available channels, but is also shaped by the interaction between pretraining dynamics and specific channel configurations—highlighting the importance of adaptive fine-tuning strategies under sensor constraints.

\section{Conclusion}
We investigated the effect of EMG channel reduction on silent speech synthesis through greedy elimination, exhaustive subset evaluation, and phoneme-level analysis. Our results showed that channels with high individual importance were not necessarily optimal when combined, highlighting the importance of complementary channel interactions. 
Fine-tuning pretrained models with random channel dropout consistently improved performance across 4–6 channel settings, outperforming models trained from scratch. These findings demonstrate that channel-aware training strategies can enable robust performance even with fewer EMG sensors, paving the way for more practical silent speech systems.

\section{Acknowledgements}
This work was partly supported by Institute of Information \& communications Technology Planning \& Evaluation (IITP) grant funded by the Korea government(MSIT) [No. RS2022-II220641, 50\%], [NO.RS-2021-II211343, 5\%] and the National Research Foundation of Korea(NRF) grant funded by the Korea government(MSIT) [No. RS-2024-00461617, 45\%].

\bibliographystyle{IEEEbib}
\bibliography{strings,refs}

\end{document}